  \def\L{\Lambda}
  \def\om{\omega}

 \def\P{\Psi} 

\def\imo{i}

\def\be{\begin{equation}}
\def\ee{\end{equation}}
\def\bea{\begin{eqnarray}}
\def\eea{\end{eqnarray}}

\documentclass[11pt]{article}
\usepackage[dvips]{graphicx}
\usepackage{longtable}
\usepackage{amssymb}
\begin{document}
\hyphenation{Schwar-zschild}

\centerline{\textbf{\Large Quasinormal modes of the charged  black hole }}
\centerline{\textbf{\Large in  Gauss-Bonnet gravity }}

\vspace{5mm}

\centerline{\textbf{Roman Konoplya}}
\centerline{Instituto de Fisica, Universidade de S\~{a}o Paulo,} 
\centerline{ C.P.66318, CEP 05315, S\~{a}o Paulo SP, Brazil}
\centerline{konoplya@fma.if.usp.br}
\vspace{5mm}

\begin{abstract}
The d-dimensional string generated gravity models lead to
Einstein-Maxwell equations with quadratic order correction term called the 
Gauss-Bonnet term. We calculate the quasinormal modes 
for the d-dimensional charged black hole in the framework of this model.
The quasinormal spectrum essentially depends upon the Gauss-Bonnet 
coupling parameter $\alpha$ which is related to the string scale,  
and is totally different from that for black
holes derived from Einstein action. 
In particular,  at large 
$\alpha$  the quasinormal modes are proportional to $\alpha$, 
and thus, have greater oscillation frequency and
damping rate, while as
$\alpha$ goes to zero the qusinormal modes approach their
Schwarzschild values. In contrary to Einstein theory black hole behavior, 
the damping rate of the charged GB black hole as a function of charge
does not contain a chracteristic maximum, but instead the monotonic 
falling down is observed. In addition, there have been obtained an 
asymptotic formula for large multipole numbers. 
\end{abstract}

\thispagestyle{empty}

\newpage

\section{Introduction}

Black holes in greater than four  space-time dimensions have attracted
a lot of interest within different theories. One of them is the
supersymmetric  string theory: its slope expansion yields 
corrections to the Einstein action. These corrections contain  higher
then the first powers of curvature. The quadratic term in curvature 
is the leading correcting term and can affect the graviton excitation 
spectrum near flat space. Thus one can consider the low-energy limit
of string theories as an  effective models of gravity in higher
dimensions. The full low-energy theory must solve the problem of
singularities in general relativity. The so called Gauss-Bonnet 
invariant is one of the most
promising candidates for such quadratic curvature corrections to the
Einstein action.  

In addition, Lovelock came to this action when
investigating the following question in classical theory of
gravitation: what is the most general divergence free second rank
symmetric tensor  in D-dimensional space-time, constructed form 
the metric and its derivatives up to  the second order?. 

The spherically symmetric solution describing neutral static  black
hole in Gauss-Bonnet gravity was obtained by Boulwar and Deser in \cite{boulwar-deser}.
The charged black hole solution was found by Wiltshire \cite{wiltshire1}.
Since that time, the properties of Gauss-Bonnet black holes were studied     
in a lot of papers \cite{1}-\cite{13}, 
with special interest to  black holes in anti-de
Sitter space-time as that having interpretation due to  ADS/CFT
correspondence.

As is known,  one of the most important characteristic of a  black
hole, which play a fundamental role for both classical (see \cite{kokkotas-review} for a
review) and quantum \cite{Dreyer} description is its quasinormal spectrum. 
The quasinormal modes are resonance oscillations which dominate at
late times of the black hole response to an external perturbation.
Recently a lot of papers (see for instance \cite{14} -\cite{20} and
references therein)  have been devoted to this subject 
motivated either by observational aspects or quantum aspects coming
from  Loop Quantum Gravity and ADS/CFT correspondence. 

At the same time the higher dimensional black holes have been
actively investigated recently within different extra dimensional
scenarios (see \cite{kanti} for a recent review). 
There, it is expected that the Planck scale is of order
of TeV, and thereby the quantum gravity may show itself at LHC where
it is expected to observe mini-black holes. The Gauss-Bonnet black
holes are being studied in this context as well (see \cite{7}
and references therein). In this connection, the quasinormal modes 
which are independent of the way of their excitation, and, thereby 
represent the ``footprints''  of a black hole, could give us the
¨footprints'' of the Gauss-Bonnet coupling, i.e. of low energy regime of  
quantum gravity.

Therefore we are interested here, what will happen with quasinormal
spectrum of the black hole if the  Gauss-Bonnet correction term will
be taken into consideration. To our surprise the first, and as far as
we are aware the only,  paper on this subject was written yet 15 years
ago \cite{iyer-GB}. There, some modes for few values of the slope parameter were 
evaluated when considering the scalar wave in the five  and sixth dimensional 
neutral Gauss-Bonnet black hole background. Here we are performing a 
thorough study of the  Gauss-Bonnet black hole quasinormal modes for 
both neutral and charged black hole which include the whole range 
of the slope parameter $\alpha$, and different multipole numbers $l$ and
different dimensionality of space-time $D$. In particular, we investigate
the asymptotic regime of large $\alpha$ and  $l$. It is shown
that the quasinormal behavior is totally different from that for the 
ordinary black hole corresponding to Einstein action. Thus, the
quasinormal behavior is crucially depend upon the slope parameter $\alpha$:
the QN modes  approach  their Schwarzschild  values  when   $\alpha$
goes to zero, and are proportional to $\alpha$ at large values of $
\alpha$. Thus, influence of Gauss-Bonnet (GB)  coupling lead to increasing of 
the oscillation frequencies and to a greater decay rate in the regime
of large GB coupling $\alpha$. For moderate values of  $\alpha$,
the the damping rate, as a function of  $\alpha$, is decreasing 
reaches some minimum, and then increasing as  $\sim \alpha$.

The dependence of the spectrum on the charge 
of the black hole is also different from that for Reissner-Nordstrom
black hole: the imaginary part is just monotonically decreasing 
as the  charge is increased, while for charged black holes derived 
from Einstein action such as Reissner-Nordstom black hole
or its dilaton extension, one has the characteristic  maximum of the
imaginary part of $\omega$ somewhere near the 0.8 of its extremal value
of charge. 

The paper is organized as follows: in Sec.II we give some preliminary 
information on Gauss-Bonnet black hole perturbations. 
Section III deals with calculation of the quasinormal spectrum 
for Gauss-Bonnet black hole. Sec IV is devoted to summary of the
result and some future perspectives of this work.

\section{Basic formulas}

The Lagrangian for the Einstein action and the Gauss-Bonnet action in  the
D-dimensional space-time model has the form:
$$
I = \frac{1} {16 \pi G_{D} } \int d^{D} x \sqrt{-g} R
$$
\begin{equation}
+ \alpha\prime \int d^{D} x \sqrt{-g} (R_{abcd}R^{abcd} - 4 R_{cd}
R^{cd}
+ R^{2})  
\end{equation}
where $\alpha\prime$ is (positive) Gauss-Bonnet coupling constant which is related  
to the Regge slope parameter or string scale.

The black hole solution for this action  is described by the  metric:

\begin{equation}
d s^{2}= -f(r) d t^{2} + f^{-1}(r) d r^{2} +r^{2} d \Omega^{2}_{D-2}.   
\end{equation}

Here the asymptotically flat solution corresponds to the following
form of the function $f$ \cite{boulwar-deser}:

\begin{equation}
f(r) = 1+ \frac{r^{2}}{2 \alpha }- \frac{r^{2}}{2 \alpha }
\sqrt{1+\frac{8 \alpha \mu  }{r^{D-1}}}. 
\end{equation}

Here $\mu$ is the integration constant which is related to the black
hole mass, and, $\alpha$ and $\alpha\prime$ are connected by the relation:
\begin{equation}
\alpha = 16 \pi G_{D} (D-3) (D-4) \alpha\prime
\end{equation}

Note that in four space-time dimensions the Gauss-Bonnet term
is the topological invariant \cite{Zwiebach}.
The gravitational and thermodynamic properties of the above black
holes were studied in a lot of papers recently
\cite{1}, \cite{2}, \cite{9}, \cite{10}, \cite{11}, \cite{12},
\cite{13}, including the 
extension to the non-zero lambda term case. 
There are  some range of parameter $ \alpha $ and $\mu $
for which a black hole shows either thermodynamical or gravitational (
or both) instability \cite{9},  \cite{12}. Yet for $D > 6$ and $D=5$ the
Gauss-Bonnet
black hole is stable against tensor type of gravitational
perturbations \cite{argentina}, which, as was shown by Gibbons and
Hartnoll \cite{gibbons-hartnoll}
is, in a sense, the decisive type of gravitational perturbations which produce 
instability at least in the Einstein theory black holes. 
The charged Gauss-Bonnet black hole corresponds to the
following form of the function $f$ \cite{wiltshire1}:

\begin{equation}
f(r) = 1+ \frac{r^{2}}{2 \alpha }- \frac{r^{2}}{2 \alpha }
\sqrt{1+\frac{8 \alpha \mu  }{r^{D-1}} -\frac{4 \alpha Q^{2} }{2 \pi
(D-2) (D-3) r^{2 D-4}}}. 
\end{equation}
Here $Q$ is the charge of the black hole. In the latter case, there
will be timelike singularity shielded by two horizons which tend to
coincide as the charge $Q$ goes to its extremal values $Q_{ext}$.
The extremal value of charge is determined by the relation:
\begin{equation}
r_{ext}^{2 (D-3)} + \alpha  \frac{D-5}{D-3} r_{ext}^{2 (D-4)} - 
\frac{Q_{ext}^{2}}{2 \pi (D-2) (D-3)} =0
\end{equation}
where 
\begin{equation}
r_{ext}^{D-3} = - \frac{1}{3} (D-5) \mu + \left( \frac{1}{4} (D-5)^{2}
\mu^{2} +  \frac{(D-4)Q_{ext}^{2}}{2 \pi (D-2) (D-3)}\right )^{1/2}.
\end{equation}
     
To consider scalar perturbations in the background of the above
Gauss-Bonnet black hole we need to consider the test scalar field
in the GB black hole background, i.e the consider the
Klein-Gordon-Fock equartion:
\begin{equation}
\Box \Phi  = \frac{1}{\sqrt{-g}} \left(g^{ \mu \nu} \sqrt{-g} 
\Phi,_{\mu}\right),_{ \nu}  = 0.
\end{equation}
After the change of the wave function $\Phi$, and going over to the
tortoise coordinate $d r^{*} =d r/f(r)$,  the last equation can be reduced to the 
Schr\"{o}dinger wave-like form:

\be\label{Wave-like-equation}%
\left(\frac{d^2}{dr^{*2}} + \om^2 - V(r^*)\right)\Psi = 0. \ee%

An effective potential has the form
\begin{equation}
V(r)=f(r)\left(\frac{(D-2)(D-4)}{4 r^{2}}f(r)+ \frac{D-2}{2r}
f'(r)+ \frac{l (l+D-3)}{r^{2}}\right)
\end{equation}

The tortoise coordinate $r^{*}$ is defined on the interval $(- \infty,  +
\infty)$ in such a way that the spatial infinity
$r=+\infty $ corresponds to $r^{*} =\infty$, while the event horizon
corresponds to   $r^{*} =-\infty$. 
The above effective potential is positively defined  and has the form of 
the potential barrier which approaches constant values at both spatial
infinity and event horizon.

\section{Quasinormal modes of GB black hole }

Under the choice of the positive sign of the real part of
$\omega$  ($\omega = Re \omega -Im \omega$), QNMs satisfy the following boundary conditions
\be\label{bounds} \P(r^*) \sim C_\pm \exp(\pm\imo\om r^*), \qquad
r\longrightarrow \pm\infty,
\ee%
corresponding to purely in-going waves at the event horizon and
purely out-going waves at infinity.

To find the quasinormal modes of the black hole whose the effective
potential has the form of potential barrier like that of the  formula (9), 
one can use the WKB approach. The latter was first used for
calculations of quasinormal modes in the first order by Schutz and
Will \cite{schutz-will}, and extended by Iyer and Will to the third order beyond the
eikonal approximation \cite{IyerWill}, and then extended to sixth
WKB order in \cite{KonoplyaWKB6}.
The WKB approach for finding QN modes is used in a lot of papers, 
and in particular the WKB corrections up to 6th order 
have been used recently in \cite{cardoso-acustic}, \cite{16}, \cite{konoplya03-3}.
The WKB formula has the form \cite{KonoplyaWKB6}:
 
\be\label{WKB}
\imo\frac{\om^2 - V_0}{\sqrt{-2V_0^{\prime\prime}}} - \L_2 - \L_3 - \L_4 -
\L_5 - \L_6 = n + \frac{1}{2},
\ee
where $V_0$ is the height and $V_0^{\prime\prime}$ is the second
derivative with respect to the tortoise coordinate of the
potential at the maximum. $\L_2$ and $\L_3$ can be found in
\cite{IyerWill}, $\L_4$, $\L_5$ and $\L_6$ are presented in
\cite{KonoplyaWKB6}; the corrections depend on the value of the potential
and higher derivatives of it at the maximum.
Below we shall discuss the obtained qusinormal modes classified
according to the multipole number for charged and neutral GB black hole.

From here and on we shall take $\mu=1$ for any dimension $D$, 
thus, in different space-time dimensions the units are different.

\subsection{The neutral black hole,  l=0}

Within the WKB approach, we shall use here, $l=0$ modes represent  an
exclusive case, 
since the unmodified WKB method is a good approximation when $l > n $, being not
so good for $l= n$ (and in fact being inapplicable when $l > n$).
As is known from comparison with numerical computations \cite{IyerWill}, 
the error at $l=0$ scalar modes can reach even 10 per cent for the
Schwarzshild black hole case. So one should be careful when
interpreting the WKB results for this case. The  $l=0$ modes
for $d=5, 6, 7, 8$-dimensional black holes are represented in the
following tables for different values of $\alpha$ :

\begin{minipage}[c]{.45\textwidth}
\begin{longtable}{|c|c|c|}
  \hline
  $\alpha$ ($D=5$) & Re $\omega$ & -Im $\omega$ \\ \hline
\hline
  $1/10$ &0.389935 & 0.256159\\ \hline
  $1/5 $ &0.396034 & 0.250548\\ \hline
  $1/2$  &0.429741 & 0.208293\\ \hline
  $5$    & --- & ---\\ \hline
  $10$   & --- & ---\\ \hline
  $20$   & --- & ---\\ \hline
\hline  
$\alpha$ ($D=6$) & Re $\omega$ & -Im $\omega$ \\ \hline
\hline
  $1/10$&0.735854 & 0.402416\\ \hline
  $1/5$&0.748053 & 0.391049 \\ \hline
  $1/2$&0.83530  & 0.304837 \\ \hline
  $5$  &0.906661 & 0.144820\\ \hline
  $10$ &1.513624 & 0.456935 \\ \hline
  $20$ &2.961675 & 0.927128 \\ \hline
\end{longtable}
\end{minipage}
\begin{minipage}[c]{.45\textwidth}
\begin{longtable}{|c|c|c|}
\hline
  $\alpha$ ($D=7$) & Re $\omega$ & -Im $\omega$ \\ \hline
\hline
  $1/10$ &1.11738  & 0.546056 \\ \hline
  $1/5$  &1.13699  & 0.529543  \\ \hline
  $1/2$  &1.29469  & 0.395111 \\ \hline
  $5$    &1.39823  & 0.574472 \\ \hline
  $10$   &1.48053  & 0.385616 \\ \hline
  $20$   &2.00368  & 0.56458 \\ \hline
\hline
  $\alpha$ ($D=8$) & Re $\omega$ & -Im $\omega$ \\ \hline
\hline
  $1/10$&1.51702 & 0.694245 \\ \hline
  $1/5$ &1.54463 & 0.673566 \\ \hline
  $1/2$ &1.7854  & 0.488086\\ \hline
  $5$   &1.47941 & 0.868859\\ \hline
  $10$  &1.800   & 0.410252\\ \hline
  $20$  &2.15426 & 0.556954\\ \hline
 \end{longtable}
\end{minipage}
$$ Table \quad  1 \quad (L=0)$$

\vspace{5mm}

We presented here a fundamental ($n=0$) overtone which is dominating in a
signal. Some discussions of higher overtone behavior will be given 
in the following subsections. Note that in five space-time diamnsions
we have a black hole-type solution only for $\alpha < 1.9$, that is why
the corresponding boxes in the table are empty.

\subsection{The neutral black hole, $ l=1, 2$ and asymptotically high $l$}

The case of higher than zero multipoles is favorable for WKB
methods. Here we observe good  convergence of the QN values when increasing the WKB order.
The $l=1, 2$ fundamental QN modes are given in the tables 2 and 3 respectively.

\begin{minipage}[c]{.45\textwidth}
\begin{longtable}{|c|c|c|}
  \hline
 $\alpha$ ($D=5$) & Re $\omega$ & -Im $\omega$ \\ \hline
\hline
  $1/10$ &0.720423  & 0.255506\\ \hline
  $1/5$  &0.723177  & 0.252684\\ \hline
  $1/2$  &0.739205  & 0.236885\\ \hline
  $5$ & --- & ---\\ \hline
  $10$ & ---  & ---\\ \hline
  $20$ & --- & ---\\ \hline
\hline
 $\alpha$ ($D=6$) & Re $\omega$ & -Im $\omega$ \\ \hline
\hline
  $1/10$ &1.139007  & 0.415034 \\ \hline
  $1/5$  &1.136193  & 0.415706   \\ \hline
  $1/2$  &1.158635  & 0.391339\\ \hline
  $5$    &1.790103  & 0.258382\\ \hline
  $10$   &3.260415  & 0.498426\\ \hline
  $20$   &6.437139  & 0.991374\\ \hline
\end{longtable}
\end{minipage}
\begin{minipage}[c]{.45\textwidth}
\begin{longtable}{|c|c|c|}
\hline
  $\alpha$ ($D=7$) & Re $\omega$ & -Im $\omega$ \\ \hline
\hline
  $1/10$ &1.54573  & 0.577608\\ \hline
  $1/5$  &1.53194  & 0.58701 \\ \hline
  $1/2$  &1.56277  & 0.554551\\ \hline
  $5$    &2.01379  & 0.308316\\ \hline
  $10$   &2.47824  & 0.423737\\ \hline
  $20$   &3.39094  & 0.59452\\ \hline
\hline
  $\alpha$ ($D=8$) & Re $\omega$ & -Im $\omega$ \\ \hline
\hline
  $1/10$& 1.93407 & 0.750046\\ \hline
  $1/5$ & 1.90391 & 0.773947\\ \hline
  $1/2$ & 1.94463 & 0.734591\\ \hline
  $5$   & 2.44511 & 0.455166\\ \hline
  $10$  & 2.6644  & 0.463966\\ \hline
  $20$  & 3.2276  & 0.581024\\ \hline
 \end{longtable}
\end{minipage}
$$ Table \quad  2 \quad (L=1)$$

\begin{minipage}[c]{.45\textwidth}
\begin{longtable}{|c|c|c|}
  \hline
  $\alpha$ ($D=5$) & Re $\omega$ & -Im $\omega$ \\ \hline
\hline
  $1/10$& 1.07492 & 0.248021  \\ \hline
  $1/5$ & 1.08169 & 0.243352  \\ \hline
  $1/2$ & 1.10459 & 0.228154 \\ \hline
  $5$   & --- & ---\\ \hline
  $10$  & --- & ---\\ \hline
  $20$  & --- & ---\\ \hline
\hline
  $\alpha$ ($D=6$) & Re $\omega$ & -Im $\omega$ \\ \hline
\hline
  $1/10$& 1.60284  & 0.39205    \\ \hline
  $1/5$ & 1.607508 & 0.387098   \\ \hline
  $1/2$ & 1.62621  & 0.370114   \\ \hline
  $5$   & 2.619279 & 0.269015   \\ \hline
  $10$  & 4.81752  & 0.505708   \\ \hline
  $20$  & 9.52384  & 1.00306    \\ \hline
\end{longtable}
\end{minipage}
\begin{minipage}[c]{.45\textwidth}
\begin{longtable}{|c|c|c|}
\hline
  $\alpha$ ($D=7$) & Re $\omega$ & -Im $\omega$ \\ \hline
\hline
  $1/10$& 2.10314  & 0.525815     \\ \hline
  $1/5$ & 2.10106  & 0.524304     \\ \hline
  $1/2$ & 2.10473  & 0.51385      \\ \hline
  $5$   & 2.66996  & 0.330935     \\ \hline
  $10$  & 3.41504  & 0.433789     \\ \hline
  $20$  & 4.69346  & 0.603341     \\ \hline
\hline
  $\alpha$ ($D=8$) & Re $\omega$ & -Im $\omega$ \\ \hline
\hline
  $1/10$& 2.58782  & 0.651764     \\ \hline
  $1/5$ & 2.57296  & 0.658318     \\ \hline
  $1/2$ & 2.54771  & 0.665021     \\ \hline
  $5$   & 3.06761  & 0.428616     \\ \hline
  $10$  & 3.49737  & 0.47698      \\ \hline
  $20$  & 4.26009  & 0.589931     \\ \hline
 \end{longtable}
\end{minipage}

$$ Table \quad  3 \quad (L=2) $$

\vspace{5mm}

\begin{figure}
\begin{center}
\includegraphics{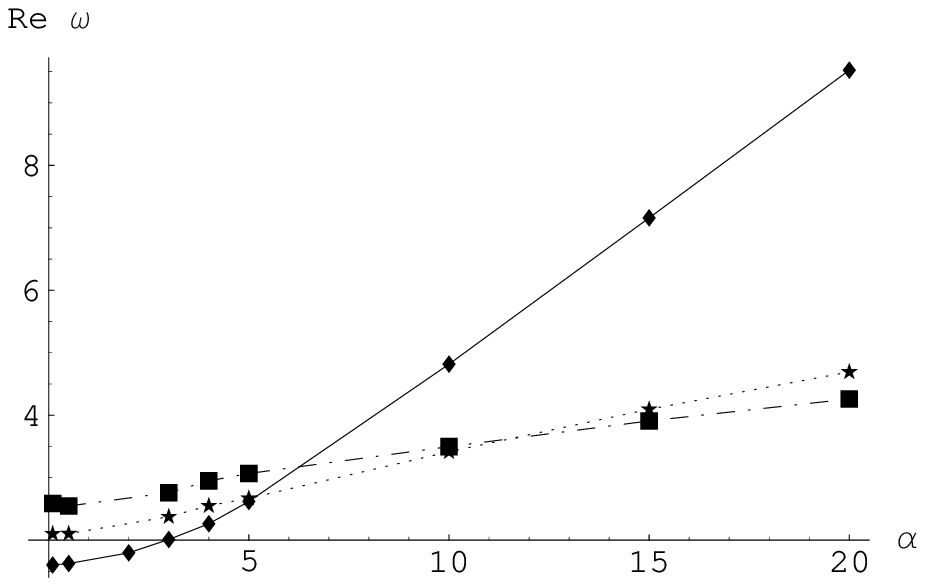}
\caption{Real part of $\omega$ as a function of the Gauss-Bonnet
coupling constant  $\alpha$ for $l=2$, $n=0$ modes of six(diamond),
seven(star) and
eight(box) dimensional black holes}
\label{1}
\end{center}
\end{figure}

\begin{figure}
\begin{center}
\includegraphics{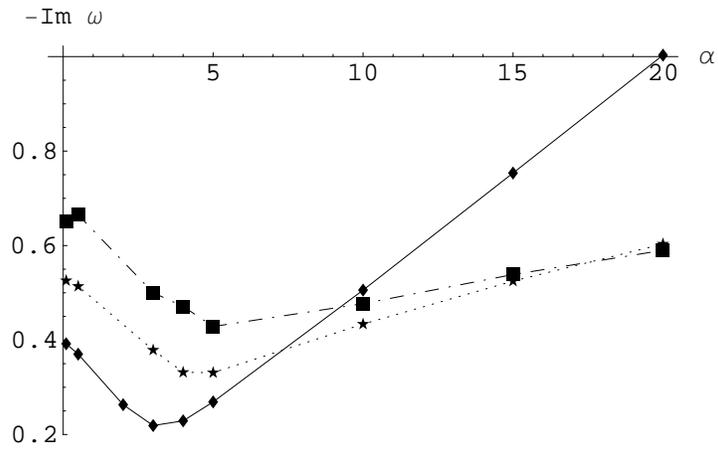}
\caption{Imaginary part of $\omega$ as a function of the Gauss-Bonnet
coupling constant  $\alpha$ for $l=2$, $n=0$ modes of six(diamond), seven(star) and
eight(box) dimensional black holes}
\label{2}
\end{center}
\end{figure}

For large multipole number $l$ we can obtain the asymptotic formula
using the first order WKB approach. For this to perform we first
represent the value of $r$ at which the effective potential attains 
the maximum in the form:
\begin{equation}
r_{max} = r_{0} + r_{1}(1/l) + r_{2} (1/l)^{2}+...   
\end{equation} 
Then we  expand the effective potential and its second  derivative in orders $1/l$ 
and insert this exapnsded values into the first order WKB formula. 
As a result for the most slow damping modes  we obtain the asymptotic formula:
\begin{equation}
\omega_{Re} =\left (\frac{1}{r_{0}^{2}}   -\frac{-1 +\sqrt{1+ 8r_{0}^{1-D}
\alpha M}}{2 \alpha} \right)  l + A, \quad \omega_{Im} = B  
\end{equation}  
where $A$ and $B$ are constants which does not depend on $l$ but
depends only on black hole parameters and  determined by cumbersome
relations and therefore are not given here.

To show that as $\alpha$ goes to zero the QN modes of GB black hole
goes to their Schwarzschild values, let us look at 
the fit for some sample mode ($D=5$, $l=2$, $n=0$), built 
on five values of $\alpha$, $\alpha$ $=$ $1/100$, $1/200$, ...$1/500$:

$$ \omega = 1.0681 - 0.2528 i +$$ 
\begin{equation}  
(0.0422 + 0.0055 i) \alpha + (2.4021 +3.8373 i) \alpha^2 +
O[\alpha^{3}].
\end{equation}   
We see that the fit is  approching the corresponding Schwarzschild
value $1.0681 - 0.2529 i$.

(Note that in \cite{konoplya03-3} the tensor type
gravitational modes are given which coincide with scalar field modes
found in \cite{KonoplyaWKB6}).  

This result can be easily explained by the fact that for small
$\alpha$ 
one has an asymptotic
$$ f(r) = 1- \frac{2 M}{r^{D-3}} + \frac{4 \alpha M}{r^{2 D -4}}+O[\alpha^{2}],
\quad \alpha \rightarrow 0, $$
and therefore as $\alpha$ goes to zero the metric approaches the Schwarzschild metric.

The dependence of the quasinormal modes on $\alpha$ can be seen from
sample figures 1 and 2 and all the above data. Namely we see that when $\alpha$ is  vanishing, both real
and imaginary parts approach its higher
dimensional Schwarzschild values, while when $\alpha$ is
becoming large, the quasinormal frequencies are proportional to
$\alpha$. Apparently it is possible to obtain some analytic formula
for 
this regime of large $\alpha$. Note that one can observe the
convergence of the results obtained with the WKB formula when increasing
the WKB order for both large and very small $\alpha$. The worst
convergence occurs for intermediate values of $\alpha$, yet this
happens only for zero multipole modes when $l=n$. For $l>n$ the 
convergence is good for the whole range of parameters.
Note that the WKB method in the Will-Schutz form uses the series which is
convergent only asymptotically, therefore we are unable 
estimate the error strictly within used method. 
Yet, in principle, it can be modified to control the error
\cite{Galtsov}. All previous expiriance in usage of the Will-Schutz
approach, based on comparison with accurate numerical results, says 
that if we observe the convergence, then it is the convergence to 
some true QN value, and the greater the WKB order the better the 
accuracy. Certainly we do not have any garantee that we shall observe 
the convergence in any case we wish.

Thus from (12) we see that for the GB black hole, similar to 
the higher dimensional Schwarzschild black hole behavior,  
for asymptotically large multipole $l$, the real part is
proportional to $l$ while the imaginary part approaches some constant.
The higher overtones within WKB method used here can be obtained only
with $l>n$. For instance, from the following data for sufficiently
large multipole ($l=5$) we can reproduced first several overtones:

$$4.73161 - 0.34419 i \quad n=0$$
$$4.70975 - 1.02805 i \quad n=1$$
$$4.64921 - 1.69859 i \quad n=2$$
$$4.49639 - 2.35764 i \quad n=3$$
$$4.13195 - 3.05609 i \quad n=4$$

In the above data  $\alpha=5$, and $D=7$.  On this example data
we see that the first few higher overtones have decreasing real part
as the imaginary part increases. That feature is similar to the
Schwarzschild black hole quasinormal behavior. To find sufficiently
high overtones one should use the continued fraction technique.

\subsection{The charged black hole}

The dependence of the quasinormal spectrum on the charge of the GB
black hole is quite different from the case of the Einstein black
holes. In particular, from \cite{iyerRN}, \cite{mePLB}, \cite{20} we know 
that the imaginary part of the quasinormal mode for Reissner-Nordstrem
or  dilaton black holes is growing as the charge is increasing until 
some its maximum value,
then, usually at the charge equaled approximately to  $0.8 Q$ there
is a sharp fall down. As can be seen from the Fig. 3, there is
no such a maximum value of the damping rate, but instead, the
monotonic decreasing exists. On the Fig.4 one can see that the real
oscillation frequency is just monotonically increases as the charge of
the black hole is increased.

\begin{figure}
\begin{center}
\includegraphics{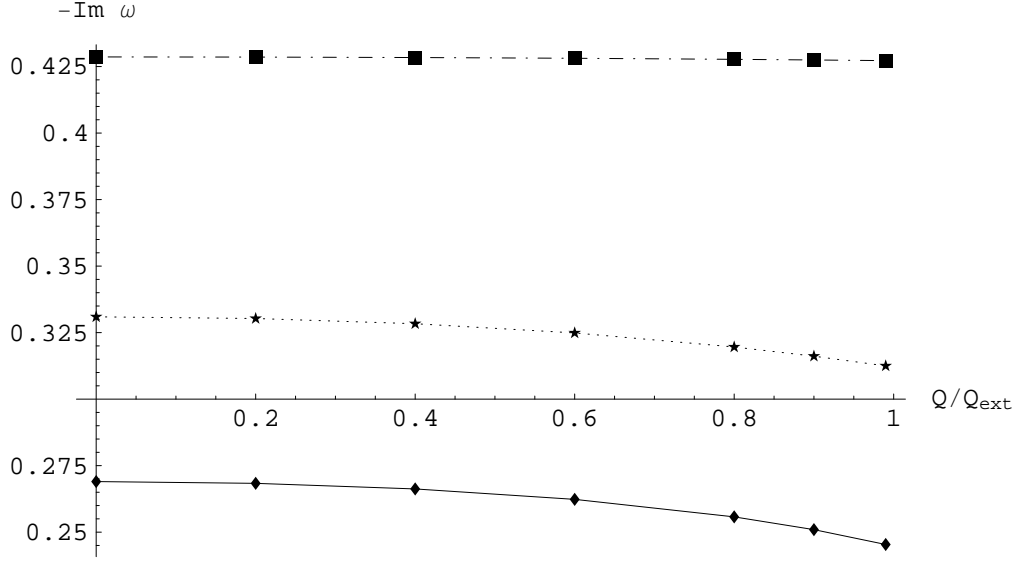}
\caption{Imaginary part of $\omega$ for D=6(diamond), 7(star), 8(box) dimensional black holes 
on example of $L=2$, $n=0$, $\alpha=5$.}
\label{3}
\end{center}
\end{figure}

\begin{figure}
\begin{center}
\includegraphics{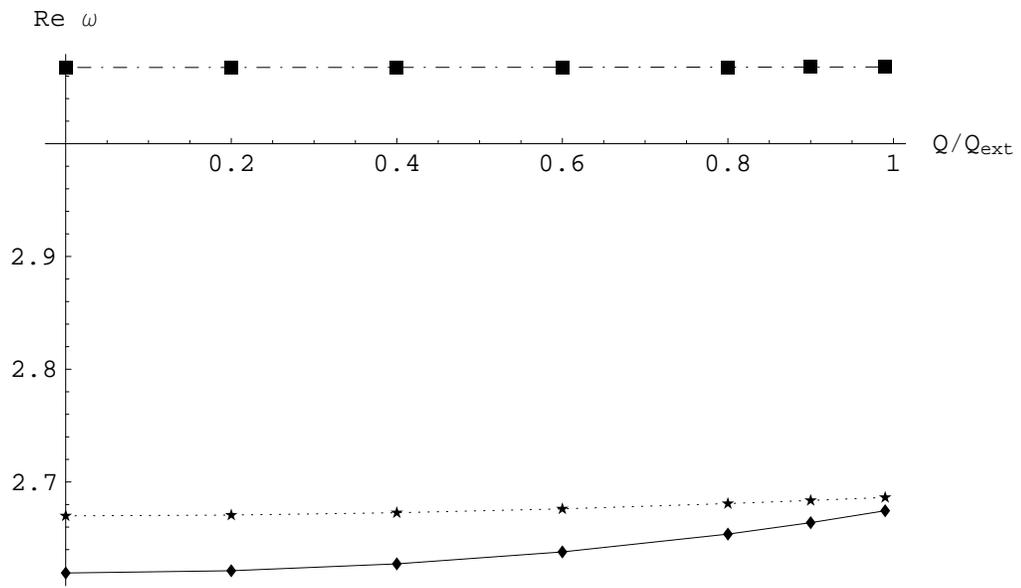}
\caption{Real part of $\omega$ for D=6(diamond), 7(star), 8(box) dimensional black holes 
on example of $L=2$, $n=0$, $\alpha=5$}
\label{4}
\end{center}
\end{figure}

At high $l$ and non extremal value of the black hole charge, in a
similar fashion to the case of the neutral black hole,  we 
obtain the asymptotic formula:
\begin{equation}
\omega_{Re} =\left(\frac{1}{r_{0}^{2}} - \frac{-1 +\sqrt{1+ 8r_{0}^{1-D}
\alpha M- \frac{2 Q^{2 } r_{0}^{4-2 D} \alpha}{6 - 5 D + D^{2}}}}{2
\alpha} \right)^{1/2} l + A', \quad \omega_{Im} =B',  
\end{equation}  
where the constants $A'$ and $B'$ reduces to  $A$ and $B$ when $Q=0$.
The explicit form of these constants and of the coefficient $r_{0}$,
$r_{1}$..etc. is available form the author upon request.

\section{Discussions}

We have investigated quasinormal modes of a black hole in Gauss-Bonnet
gravity. It is shown that quasinormal behavior is totally different
from that of black hole in Einstein gravity. In particular, the
quasinoraml spectrum essetailly depends on the Gauss-Bonnet
coupling parameter: at large $\alpha$ the QN modes are
proportional to $\alpha$, while when $\alpha$ goes to zero, the QN
modes approach their Schwarzschild values. 
Due to the Gauss-Bonnet coupling the quasinormal modes have 
greater oscillation frequency and greater  damping rate at
large $\alpha$, but at moderate $\alpha$, the damping rate as a
function of  $\alpha$, first, is decreasing until some minimum value
and then begin to grow by the low  $\sim \alpha$. 
As can be seen from comparison of the scalar filed potential 
considered here with that found in 
\cite{12},  for tensor-type gravitational perturbations, the 
scalar field potential and the tensor type gravitational potential
coincide. The same situation takes place for higher dimensional black
holes in Einstein gravity \cite{ishibashi-kodama}. 
Thus the quasinormal spectra are the same for adequate multipoles. 
If to disentangle equations for 
other types of gravitational perturbations of GB black hole is 
a solvable problem 
(see for instance references in \cite{ishibashi-kodama}, \cite{mePLA}), then
the question of stability and gravitational QN radiation would be 
completed.

The problem that was beyond the present study is the
finding of the asymptotically high overtones of the QN spectrum.   
After all, the quasinormal modes for the GB black hole 
in Anti-de Sitter space-time may have the ADC/CFT holographic 
interpretation and are also interesting to study. We hope that
further investigation will clarify these points.

\section{Acknowledgements}
The author would like to thank Ishware Neupanee for useful discussions.  
This work was supported by FAPESP (Brazil).


\end{document}